# LOCALIZATION OF $M$ -PARTICLE QUANTUM WALKS

**CLEMENT AMPADU**

31 Carrolton Road
Boston, Massachusetts, 02132
U.S.A.
e-mail: drampadu@hotmail.com

**Abstract**

We study the motion of $M$ particles performing a quantum walk on the line. Under various conditions on the initial coin states for quantum walkers controlled by the Hadamard operator, we give theoretical criterion to observe the quantum walkers at an initial location with high probability.

**KEY WORDS:** quantum walks, Hadamard operator, localization

**PACS:** 03.67.-a

## I. INTRODUCTION

In this paper we study the localization problem on the line for $M$ walkers controlled by the Hadamard operator given at the end of Sec. II of this paper. In particular, we are interested in the answer to the following question: If, say, the $M$ walkers exists at one site initially on the line, will the quantum walkers remain trapped with high probability? The answer to this question has been investigated by various authors in varying contexts under the pseudonym "localization". This paper to the author's knowledge is the first to study localization problem for a general $M$ -particle walk, whilst the previous studies focused on the one-particle quantum walk. For example, Inui et.al [1] studied a generalized Hadamard walk in one dimension with three inner states and concluded that the quantum walker (quantum particle) is trapped near the origin with high probability. Watabe et.al [2], on the other hand, were able to control localization around the origin for a one-parameter family of discrete-time quantum walk models on the square lattice, which included the Grover walk, which is related to the Grover's algorithm in computer science, as a special case. For the Grover walk itself in two dimensions,

Inui et.al [3] where able to show localization analytically. Liu and Pentulante [4] were able to offer theoretical explanation for localization in the case of discrete random walks on a linear lattice with two entangled coins. Recently, Ampadu [5] studied a one parameter family of discrete-time quantum walk models on $Z$ and $Z^2$ associated with the Hadamard walk and gave necessary conditions for the existence of localization. Following the localization studies by these and many other authors in the literature, we give theoretical criterion for the $M-$particle quantum walk controlled by the Hadamard operator given at the end of Sec. II of this paper.

This paper is organized as follows, in Sec.II we briefly review the concept of the quantum walk on the line with $M$ particles, in Sec. III and Sec. IV we generalize the $M-$particle walk for both distinguishable and indistinguishable particles and give the probability distribution generated by the $M-$particle walk in both cases. In Sec.V we give the theoretical criterion for observing the quantum walkers at an initial location with high probability. Since in an $M-$particle quantum walk the walkers are at different positions on the line, we first solve the meeting problem [6], which ask for the probability that the $M$ walkers will be detected at the same position after $t$ steps. Next, we give the the probability of localization anywhere on the line for distinguishable and indistinguishable particles. For distinguishable particles, these are given by Theorem 5.1 for factorized initial states and Theorem 5.2 for entangled initial states. For indistinguishable particles, which we only state for bosons, these are given by Theorems 5.4 for probability amplitudes that do not factorize, and Theorem 5.5 for probability amplitudes that do factorize. The proofs of these theorems is essentially the same as given by Theorem 4.1 in Ampadu[5], thus we omit them. Note that the localization probability is given in terms of the intensity of the Dirac delta function, by making use of the stationary probability distribution $P(m) \equiv \lim_{t\to\infty} \sum_{m_1} P(m_1,\cdots,m_1;t)$ . In particular, $P(m_1,\cdots,m_1;t)$, only gives us the probability that the other particles meets the first at its position at $m_1$ after $t$ steps, in contrast to $\sum_{m_1} P(m_1,\cdots m_1;t)$ which gives the probability of the particles meeting anywhere on the line at time $t$.

Finally we state our localization criterion in Definitions 5.3 for distinguishable particles, and for indistinguishable particles in Definition 5.6 for the case of bosons. Sec. VI is devoted to the conclusions.

## II. QUANTUM WALK ON THE LINE WITH $M$ PARTICLES

Consider a quantum walk on the line involving $M$ particles. Let $H$ be the Hilbert space of the particles which consist of the position space $H_p$ spanned by the set of orthonormal states $\{i \mid i \in Z\}$, and the coin space $H_C$ spanned by $\{|L\rangle, |R\rangle\}$. The Hilbert space of the $M$ particles is given by the tensor product of the single walker spaces, say $H = (H_p \otimes H_c)_1 \otimes (H_p \otimes H_c)_2 \otimes \cdots \otimes (H_p \otimes H_c)_M$. To define the movement of the particles in one dimension, we first consider what happens on one step in the quantum walk. For each of the $M$ particles, we first make superposition on the coin space with coin operator $U_C$ and move the particle according to the coin state with the translation operator $S = \sum_x \{|x-1\rangle\langle x| \otimes |L\rangle\langle L| + |x+1\rangle\langle x| \otimes |R\rangle\langle R|\}$, the evolution operator for the $M-$particle quantum walk is then given by $U_W = (S\cdot(I \otimes U_C))_1 \otimes \cdots \otimes (S\cdot(I \otimes U_C))_M$, where $I$ is the identity operator in the single particle's position space, and $U_W$ is the coin operator on the position space of the $M$ particles combined. The evolution of the quantum walk is then defined by $|\Psi(t+1)\rangle = U_W|\Psi(t)\rangle$, which by induction on $t$, can be written in terms of the initial state of the particle as $|\Psi(t+1)\rangle = U_W^t|\Psi(0)\rangle$. Note that in the above description of the quantum walk on the line it is assumed that interaction between the walkers does not exist, that is, it is assumed the walkers are distinguishable, otherwise the evolution operator does not factorize which corresponds to the case where the walkers are indistinguishable. In this paper we will take $U_C = H^*\left(\frac{1}{2},\frac{1}{2}\right)$, where $H^*(p,q)$ is a one dimensional generalization of the Hadmard walk, Ampadu [5], for example.

## III. DISTINGUISHABLE PARTICLES

Recall in the case that there is no interaction between the walkers, the Hilbert space of the $M$ walkers is given by $H = (H_p \otimes H_c)_1 \otimes (H_p \otimes H_c)_2 \otimes \cdots \otimes (H_p \otimes H_c)_M$. The wave function of the walkers at time $t$ describing the state of the system is given by the vectors

$$\psi(m_1, m_2, \cdots, m_M; t) = \begin{pmatrix} \psi_{LL\cdots L}(m_1, \cdots, m_M; t) \\ \psi_{LR\cdots LR}(m_1, \cdots, m_M; t) \\ \psi_{RL\cdots RL}(m_1, \cdots, m_M; t) \\ \psi_{LR\cdots RL}(m_1, \cdots, m_M; t) \\ \psi_{RL\cdots LR}(m_1, \cdots, m_M; t) \\ \psi_{RR\cdots R}(m_1, \cdots, m_M; t) \end{pmatrix}$$

. Note that this description is for a general $M-$particle walk. For example the component $\psi_{LL\cdots L}(m_1, \cdots, m_M; t)$ is the amplitude of the state where the walkers are on sites $m_1, \cdots, m_M$ respectively with the internal state $|L\rangle$ for each of them. The state of the walkers at time $t$ is then given by

$$|\psi(t)\rangle = \sum_{m_1, \cdots, m_M} \sum_{k_1, \cdots, k_M = L, R} \psi_{k_1, \cdots, k_M}(m_1, \cdots, m_M; t) \prod_{i=1}^{M} |m_i, k_i\rangle_i$$

. The probability that the walkers are at sites $m_1, \cdots, m_M$ at time $t$ is given by

$$P(m_1, \cdots, m_M; t) = \sum_{k_1, \cdots, k_M = L, R} \left| \prod_{i=1}^{M} \langle m_i, k_i | \psi(t)\rangle \right|^2 = \sum_{k_1, \cdots, k_M = L.R} \left| \psi_{k_1 \cdots, k_M}(m_1, \cdots, m_M; t) \right|^2$$

### A. Seperable Initial States

In the case that the $M$ walkers are initially in a factorized state say,

$$|\psi(0)\rangle = \left( \sum_{m_1, k_1} \psi_{1k_1}(m_1, 0) |m_1, k_1\rangle_1 \right) \otimes \cdots \otimes \left( \sum_{m_M, k_M} \psi_{Mk_M}(m_M, 0) |m_M, k_M\rangle_M \right)$$

, which translates into

$\psi_{k_1\cdots,k_M}(m_1,m_2,\cdots,m_M;0)=\prod_{i=1}^{M}\psi_{ik_i}(m_i,0)$, the probability distribution is a product of a single walker probability distributions which is given by $P(m_1,\cdots,m_M;t)=\prod_{i=1}^{M}P_i(m_i,t)$, where $P_i(m_i,t)=\sum_j \langle m_i,j|\rho_i(t)|m_i,j\rangle$ and $\rho_i(t)=Tr_{j\neq i}|\psi(t)\rangle\langle\psi(t)|$

**B. Entangled Initial States**

In the case that the initial state of the $M-$walkers is entangled, say,

$|\psi(0)\rangle=\sum_{\alpha}\left[\left(\sum_{m_1,k_1}\psi^{\alpha}{}_{1k_1}(m_1,0)|m_1,k_1\rangle_1\right)\otimes\cdots\otimes\left(\sum_{m_M,k_M}\psi^{\alpha}{}_{mk_M}(m_M,0)|m_M,k_M\rangle_M\right)\right]$, the probability distribution cannot be expressed in terms of single walker distributions, but in terms of probability amplitudes as $P(m_1,m_2,\cdots,m_M;t)=\sum_{k_1,\cdots,k_M=L,R}\left|\sum_{\alpha}\left(\prod_{i=1}^{M}\psi^{\alpha}_{ik_i}(m_i,t)\right)\right|^2$.

*Example (Bell-Type Basis)*

In the case of initial states of the form $|\psi^{\pm}\rangle=\frac{1}{\sqrt{2}}(|LR\cdots LR\rangle\pm|RL\cdots RL\rangle\pm|LR\cdots RL\rangle\pm|RL\cdots LR\rangle)$ and $|\phi^{\pm}\rangle=\frac{1}{\sqrt{2}}(|LL\ldots L\rangle\pm|RR\ldots R\rangle)$, Ampadu [7] has shown that the joint probability distribution generated by the $M-$particle quantum walk with the initially entangled coins described by the Bell-type basis, is given by

$$P^{(\psi^{\pm})}(m_1,m_2,\ldots,m_M,t)=\frac{1}{2}\sum_{k_1,\ldots,k_M=L,R}\left|\begin{array}{l}\psi^{(LR\ldots LR)}_{k_1,k_2,\ldots,k_M}(m_1,m_2,\ldots,m_M,t)\pm\psi^{(RL\ldots RL)}_{k_1,k_2,\ldots,k_M}(m_1,m_2,\ldots,m_M,t)\pm\\ \psi^{(LR\ldots RL)}_{k_1,k_2,\ldots,k_M}(m_1,m_2,\ldots,m_M,t)\pm\psi^{(RL\ldots LR)}_{k_1,k_2,\ldots,k_M}(m_1,m_2,\ldots,m_M,t)\end{array}\right|^2$$

and

$$P^{(\phi^{\pm})}(m_1,m_2,\ldots,m_M,t)=\frac{1}{2}\sum_{k_1,\ldots,k_M=L,R}\left|\psi_{k_1,k_2,\ldots,k_M}^{(LL\ldots LL)}(m_1,m_2,\ldots,m_M,t)\pm\psi_{k_1,k_2,\ldots,k_M}^{(RR\ldots RR)}(m_1,m_2,\ldots,m_M,t)\right|^2$$

where the superscript in both the formulas $P^{(\psi^{\pm})}(m_1,m_2,\ldots,m_M,t)$ and $P^{(\phi^{\pm})}(m_1,m_2,\ldots,m_M,t)$ indicates the initial coin state.

## IV. INDISTINGUISHABLE PARTICLES

In the case that the $M$ walkers are indistinguishable, it is natural to use the so called second quantization formalism, in particular the time evolution is now given by the transformation of creation operators- bosons and fermions (see Ampadu[7] or Stefanak et.al [6,8] for details). In particular the state of the $M$ bosonic and fermionic walkers are given by

$$\left|\psi^{Bosonic}(t)\right\rangle=\sum_{m_1,\cdots,m_M}\sum_{r_1,\cdots r_{M=L,R}}\frac{1}{2}\left(\psi_{r_1r_2\cdots r_{M-1}r_M}(m_1,m_2,\cdots,m_M;t)+\psi_{r_Mr_{M-1}\cdots r_2r_1}(m_M,m_{M-1},\cdots,m_1;t)\right)\prod_{i=1}^{M}{}^{T}\,\bar{\hat{a}}_{(m_i,r_i)}\left|vac\right\rangle$$

$$\left|\psi^{Fermionic}(t)\right\rangle=\sum_{m_1,\cdots,m_M}\sum_{r_1,\cdots r_{M=L,R}}\frac{1}{2}\left(\psi_{r_1r_2\cdots r_{M-1}r_M}(m_1,m_2,\cdots,m_M;t)-\psi_{r_Mr_{M-1}\cdots r_2r_1}(m_M,m_{M-1},\cdots,m_1;t)\right)\prod_{i=1}^{M}{}^{T}\,\bar{\hat{b}}_{(m_i,r_i)}\left|vac\right\rangle$$

, where $\left|vac\right\rangle$ denotes the vacuum state. For $m_i\geq m_{i+1}$ and $m_i\neq m_{i+1}$ the probability distribution is given by

$$P^{Bosonic,Fermionic}(m_1,m_2,\cdots,m_M;t)=\sum_{r_1,\cdots,r_M=L,R}\left|\left\langle\prod_{i=1}^{M}1_{(m_i,r_i)}\middle|\psi^{Bosonic,Fermionic}(t)\right\rangle\right|^2=\sum_{r_1,r_2,\cdots,r_M=L,R}\left|\psi_{r_1r_2\cdots r_{M-1}r_M}(m_1,\cdots,m_M;t)\pm\psi_{r_Mr_{M-1}\cdots r_2r_1}(m_M,\cdots,m_1;t)\right|^2$$

where the plus sign on the right corresponds to bosonic and the negative sign corresponds to fermionic.

*Example IV.1 (Probability Amplitudes Factorize)*

Assume $\psi_{r_1\cdots r_M}(m_1,m_2,\cdots,m_M;t)=\prod_{i=1}^{M}\psi_{ir_i}(m_i,t)$ and suppose that the initial state of the coin is given by $\left|1_{0,\cdots0;LR\cdots LR}\right\rangle$, Ampadu [7] have shown that

$$P^{Bosonic,Fermionic}(m_1, m_2, m_3, \cdots, m_M; t)$$
$$= \sum_{r_1,\cdots,r_M=L,R} \left| \Psi^{LR\cdots LR}_{r_1\cdots r_M}(m_1, m_2, \cdots, m_M, t) \pm \Psi^{RL\cdots RL}_{r_1\cdots r_M}(m_1, m_2, \cdots, m_M, t) \pm \Psi^{LR\cdots RL}_{r_1\cdots r_M}(m_1, m_2, \cdots, m_M, t) \pm \Psi^{RL\cdots LR}_{r_1\cdots r_M}(m_1, m_2, \cdots, m_M, t) \right|^2$$

where the plus sign on the right corresponds to bosonic and the negative sign corresponds to fermionic.

*Example IV.2 (Bell-Type Basis)*

For initial states described by the example in Section *III.B*, and the example immediately above, Ampadu [7] have shown the following relations $P^{Bosons}(m_1.\cdots, m_M; t) = 2P^{\psi^+}(m_1, m_2, \cdots, m_M; t)$ and $P^{Fermions}(m_1.\cdots, m_M; t) = 2P^{\psi^-}(m_1, m_2, \cdots, m_M; t)$.

## V. THEORETICAL CRITERION FOR LOCALIZATION

### A. Distinguishable Particles

In this case the general meeting probability is given by the norm of the vector $\psi(m, \cdots, m; t)$ $P(m, \cdots, m; t) = \sum_{k_1,\cdots,k_M=L,R} \psi_{k_1\cdots k_M}(m, \cdots, m; t)$. When the $M$-walkers are initially in a factorized state as in Section *III.A*, then the meeting probability can be further simplified to the multiple of the probabilities that the individual walkers will reach the site, in particular, $P(m_1, \cdots, m_1; t) = [P_1(m_1, t)]^M$ where $P_1(m_1, t) = \sum_{\substack{m_i \\ i \neq 1}} P(m_1, \cdots, m_M; t)$. On the other hand when the $M$-walkers are in an entangled state as in Section *III.B*, then the meeting probability is given by $P(m_1, \cdots, m_1; t) = \sum_{k_1,\cdots,k_M=L,R} \left| \sum_{\alpha} \left[\psi^{\alpha}_{1k_1}(m_1, t)\right]^M \right|^2$.

*Theorem 5.1 (Localization Probability for Factorized Initial States):* Define

$P(m) \equiv \lim_{t\to\infty} \sum_{m_1} P(m_1,\cdots,m_1;t)$, then the probability of localization at $m = m_0$ is given by

$P(m_0) = \int_{-\infty}^{\infty} P(m)\delta(m-m_0)dm$, where $P(m_1,\cdots,m_1;t) = \left[P_1(m_1,t)\right]^M$ and

$P_1(m_1,t) = \sum_{\substack{m_i \\ i\neq 1}} P(m_1,\cdots,m_M;t)$.

*Theorem 5.2 (Localization Probability for Entangled Initial States):*

Define $P(m') \equiv \lim_{t\to\infty} \sum_{m_1} P(m_1,\cdots,m_1;t)$, then the probability of localization at $m' = y_0$ is given by

$P(y_0) = \int_{-\infty}^{\infty} P(m')\delta(m'-y_0)dm'$, where $P(m_1,\cdots,m_1;t) = \sum_{k_1,\cdots,k_M=L,R} \left| \sum_{\alpha} \left[\psi^{\alpha}_{1k_1}(m_1,t)\right]^M \right|^2$.

*Definition 5.3:* We say localization has occurred if either $P(y_0)$ or $P(m_0)$ is sufficiently large within the confines of the interval $[0,1]$

## B. Indistinguishable Particles

In this case the meeting probabilities are given, for bosons and fermions, by

$$\begin{aligned} P^{Bosonic}(m_1,\cdots,m_1;t) &= \left|\left\langle M_{(m_1,L)}\right|\psi^{Bosonic}(t)\right\rangle\right|^2 + \left|\left\langle M_{(m_1,R)}\right|\psi^{Bosonic}(t)\right\rangle\right|^2 + \left|\left\langle 1_{(m_1,L)}\cdots 1_{(m_1,R)}\right|\psi^{Bosonic}(t)\right\rangle\right|^2 \\ &= M\left|\psi_{LL\cdots L}(m_1,\ldots,m_1;t)\right|^2 + M\left|\psi_{RR\cdots R}(m_1,\ldots,m_1;t)\right|^2 + \left|\psi_{LR\cdots LR}(m_1,\cdots,m_1;t) + \psi_{RL\cdots RL}(m_1,\cdots,m_1;t) + \psi_{LR\cdots RL}(m_1,\cdots,m_1;t) + \psi_{RL\cdots LR}(m_1,\cdots,m_1;t)\right|^2 \end{aligned}$$

$$P^{Fermionic}(m_1,\cdots,m_1;t) = \left|\left\langle 1_{(m_1,L)}\cdots 1_{(m_1,R)}\right|\psi^{Fermionic}(t)\right\rangle\right|^2 = \left|\psi_{LR\cdots LR}(m_1,\cdots,m_1;t) - \psi_{RL\cdots RL}(m_1,\cdots,m_1;t) - \psi_{LR\cdots RL}(m_1,\cdots,m_1;t) - \psi_{RL\cdots LR}(m_1,\cdots,m_1;t)\right|^2$$

In the case that the probability amplitudes factorize, say can write

$\psi_{r_1\cdots r_M}(m_1,m_2,\cdots,m_M;t)=\prod_{i=1}^{M}\psi_{ir_i}(m_i,t)$, then the meeting probabilities for bosons and fermions are given by

$$P^{Bosonic}(m_1,\cdots,m_1;t)=M\left|\prod_{i=1}^{M}\psi_{iL}(m_1,t)\right|^2+M\left|\prod_{i=1}^{M}\psi_{iR}(m_1,t)\right|^2+$$
$$\left|\begin{array}{l}\psi_{1L}(m,t)\psi_{2R}(m_1,t)\cdots\psi_{(M-1)L}(m_1,t)\psi_{MR}(m_1,t)+\psi_{1R}(m,t)\psi_{2L}(m_1,t)\cdots\psi_{(M-1)R}(m_1,t)\psi_{ML}(m_1,t)+\psi_{1L}(m,t)\psi_{2R}(m_1,t)\cdots\psi_{(M-1)R}(m_1,t)\psi_{ML}(m_1,t)+\\ \psi_{1R}(m,t)\psi_{2L}(m_1,t)\cdots\psi_{(M-1)L}(m_1,t)\psi_{MR}(m_1,t)\end{array}\right|^2$$

$$P^{Fermionic}(m_1,\cdots,m_1;t)=$$
$$\left|\begin{array}{l}\psi_{1L}(m,t)\psi_{2R}(m_1,t)\cdots\psi_{(M-1)L}(m_1,t)\psi_{MR}(m_1,t)-\psi_{1R}(m,t)\psi_{2L}(m_1,t)\cdots\psi_{(M-1)R}(m_1,t)\psi_{ML}(m_1,t)-\psi_{1L}(m,t)\psi_{2R}(m_1,t)\cdots\psi_{(M-1)R}(m_1,t)\psi_{ML}(m_1,t)\\ -\psi_{1R}(m,t)\psi_{2L}(m_1,t)\cdots\psi_{(M-1)L}(m_1,t)\psi_{MR}(m_1,t)\end{array}\right|^2$$

*Theorem 5.4 (Probability Amplitudes do not factorize):* Define $P(m'')\equiv\lim_{t\to\infty}\sum_{m_1}P^{Bosonic}(m_1,\cdots,m_1;t)$, then the probability of localization at $m''=m_0'$ is given by $P(m_0')=\int_{-\infty}^{\infty}P(m'')\delta(m''-m_0')dm''$, where

$$P^{Bosonic}(m_1,\cdots,m_1;t)=\left|\left\langle M_{(m_1,L)}\middle|\psi^{Bosonic}(t)\right\rangle\right|^2+\left|\left\langle M_{(m_1,R)}\middle|\psi^{Bosonic}(t)\right\rangle\right|^2+\left|\left\langle 1_{(m_1,L)}\cdots 1_{(m_1,R)}\middle|\psi^{Bosonic}(t)\right\rangle\right|^2$$
$$=M\left|\psi_{LL\cdots L}(m_1,\ldots,m_1;t)\right|^2+M\left|\psi_{RR\cdots R}(m_1,\ldots,m_1;t)\right|^2+\left|\psi_{LR\cdots LR}(m_1,\cdots,m_1;t)+\psi_{RL\cdots RL}(m_1,\cdots,m_1;t)+\psi_{LR\cdots RL}(m_1,\cdots,m_1;t)+\psi_{RL\cdots LR}(m_1,\cdots,m_1;t)\right|^2$$

*Theorem 5.5 (Probability Amplitudes factorize):* Define $P(m''')\equiv\lim_{t\to\infty}\sum_{m_1}P^{Bosonic}(m_1,\cdots,m_1;t)$, then the probability of localization at $m'''=y_0'$ is given by $P(y_0')=\int_{-\infty}^{\infty}P(m''')\delta(m'''-y_0')dm'''$, where

$$P^{Bosonic}(m_1,\cdots,m_1;t)=M\left|\prod_{i=1}^{M}\psi_{iL}(m_1,t)\right|^2+M\left|\prod_{i=1}^{M}\psi_{iR}(m_1,t)\right|^2+$$
$$\left|\begin{array}{l}\psi_{1L}(m,t)\psi_{2R}(m_1,t)\cdots\psi_{(M-1)L}(m_1,t)\psi_{MR}(m_1,t)+\psi_{1R}(m,t)\psi_{2L}(m_1,t)\cdots\psi_{(M-1)R}(m_1,t)\psi_{ML}(m_1,t)+\psi_{1L}(m,t)\psi_{2R}(m_1,t)\cdots\psi_{(M-1)R}(m_1,t)\psi_{ML}(m_1,t)+\\ \psi_{1R}(m,t)\psi_{2L}(m_1,t)\cdots\psi_{(M-1)L}(m_1,t)\psi_{MR}(m_1,t)\end{array}\right|^2$$

*Definition 5.6:* We say localization has occurred if either $P(y_0')$ or $P(m_0')$ is sufficiently large within the confines of the interval $[0,1]$.

## VI. CONCLUDING REMARKS

In this paper we have studied the motion of $M$ particles performing a quantum walk on the line and under various conditions on the initial coin states for quantum walkers controlled by the Hadamard operator have given theoretical criterion to observe the quantum walkers at an initial location with high probability. Stefanak et.al [8] noted in their work involving directional correlations in quantum walk with two particles, that entanglement in two-particle non-interacting quantum walks cannot break the limit of probabilities they found for separable particles, and posed the following question: What happens if we consider interacting particles? This motivated them to introduce the concept of two-particle quantum walks with $\delta -$interaction to the solution of their question. The authors found out that by introducing a $\delta -$interaction one can exceed the limit derived for non-interacting particles. Recently, Ampadu[7] commenced the study of this new model of the quantum walk focusing on the Fourier analysis for the transformation $C_\delta = \frac{1}{2}\begin{pmatrix} 1 & 1 & 1 & 1 \\ 1 & -1 & -1 & 1 \\ -1 & 1 & -1 & 1 \\ -1 & -1 & 1 & 1 \end{pmatrix}$. It is an interesting problem to study the localization problem for the quantum walk with $\delta -$interaction.